# A Fast, Minimal Memory, Consistent Hash Algorithm

John Lamping, Eric Veach
Google

**Abstract**

We present jump consistent hash, a fast, minimal memory, consistent hash algorithm that can be expressed in about 5 lines of code. In comparison to the algorithm of Karger et al., jump consistent hash requires no storage, is faster, and does a better job of evenly dividing the key space among the buckets and of evenly dividing the workload when the number of buckets changes. Its main limitation is that the buckets must be numbered sequentially, which makes it more suitable for data storage applications than for distributed web caching.

**Introduction**

Karger et al. `[1]` introduced the concept of consistent hashing and gave an algorithm to implement it. Consistent hashing specifies a distribution of data among servers in such a way that servers can be added or removed without having to totally reorganize the data. It was originally proposed for web caching on the Internet, in order to address the problem that clients may not be aware of the entire set of cache servers.

Since then, consistent hashing has also seen wide use in data storage applications. Here, it addresses the problem of splitting data into a set of *shards*, where each shard is typically managed by a single server (or a small set of replicas). As the total amount of data changes, we may want to increase or decrease the number of shards. This requires moving data in order to split the data evenly among the new set of shards, and we would like to move as little data as possible while doing so.

Assume, for example, that data consisting of key-value pairs is to be split into 10 shards. A simple way to split the data is to compute a hash, h(key), of each key, and store the corresponding key-value pair in shard number h(key) mod 10. But if the amount of data grows, and now needs 12 shards to hold it, the simple approach would now assign each key to shard h(key) mod 12, which is probably not the same as h(key) mod 10; the data would need to be completely rearranged among the shards.

But it is only necessary to move 1/6 of the data stored in the 10 shards in order to end up with the data balanced among 12 shards. Consistent hashing provides this. Our jump consistent hash function takes a key and a number of buckets (i.e., shards), and returns one of the buckets. The function satisfies the two properties: (1) about the same number of keys map to each bucket, and (2) the mapping from key to bucket is perturbed as little as possible when the number of buckets is changed. Thus, the only data that needs to move when the number of buckets changes is the data for the relatively small number of keys whose bucket assignment changed.

The jump consistent hash algorithm is fast and has a large memory advantage over the one presented in Karger et al. Their algorithm needs thousands of bytes of storage per candidate shard in order to get a fairly even distribution of keys. In a large data storage application, where there may be thousands of shards, that means that each client needs megabytes of memory for its data structures, which must be stored long term for the algorithm to be efficient. In contrast, jump consistent hash needs no memory beyond what fits in a few registers. Jump consistent hash also does a better job of splitting the keys evenly among the buckets, and of splitting the rebalancing workload among the shards. On the other hand, jump consistent hash does not support arbitrary server names, but only returns a shard number; it is thus primarily suitable for the data storage case.

Figure 1 shows a complete implementation of jump consistent hash. Its inputs are a 64 bit key and the number of buckets. It outputs a bucket number in the range [0, num_buckets). The rest of this note explains what is going on in this code and gives theoretical and empirical performance results.

```
int32_t JumpConsistentHash(uint64_t key, int32_t num_buckets) {
  int64_t b = -1, j = 0;
  while (j < num_buckets) {
    b = j;
    key = key * 2862933555777941757ULL + 1;
    j = (b + 1) * (double(1LL << 31) / double((key >> 33) + 1));
  }
  return b;
}
```

Figure 1: Jump Consistent Hash algorithm in C++.

**Related work**

Karger et al.'s consistent hash algorithm works by associating each bucket with a number of randomly chosen points on the unit circle. Given a key, it hashes the key to a position on the unit circle, proceeds along the circle in a clockwise direction from there until it finds the first chosen point, and returns the bucket associated with that point. Storing the association requires memory proportional to the number of buckets times the number of points chosen per bucket. Karger et al.'s experiments used 1000 points per bucket to get to a standard deviation of 3.2% in the number of keys assigned to different buckets.

The only other algorithm we are aware of that computes a consistent hash is the rendezvous algorithm by Thaler and Ravishankar [3]. Used as a consistent hash, the original version of their algorithm takes a key, and for each candidate bucket, computes a hash function value h(key, bucket). It then returns the bucket for which the hash yielded the highest value. This requires time proportional to the number of buckets. Wang et al. [4] show how the buckets can be organized into a tree to make the time proportional to the log of the number of buckets. But their variant comes at the cost of balance when shards are added or removed, because they only re-balance across the lowest level nodes of their tree.

Both of these algorithms allow buckets to have arbitrary ids, and handle not only new buckets being added, but also arbitrary buckets being removed. This ability to add or remove buckets in any order can be valuable for cache servers where the servers are purely a performance improvement. But for data storage applications, where each bucket represents a different shard of the data, it is not acceptable for shards to simply disappear, because that shard is only place where the corresponding data is stored. Typically this is handled by either making the shards redundant (with several replicas), or being able to quickly recover a replacement, or accepting lower availability for some data. Server death thus does not cause reallocation of data; only capacity changes do. This means that shards can be assigned numerical ids in increasing order as they are added, so that the active bucket ids always fill the range [0, num_buckets).

Only two of the four properties of consistent hashing described in the Karger et al. paper are important for data storage applications. These are *balance*, which essentially states that objects are evenly distributed among buckets, and *monotonicity*, which says that when the number of buckets is increased, objects move only from old buckets to new buckets, thus doing no unnecessary rearrangement. Their other two properties, *spread* and *load*, both measure the behavior of the hash function under the assumption that each client may see a different arbitrary subset of the buckets. Under our data storage model this cannot happen, because all clients see the same set of buckets [0, num_buckets). This restriction enables jump consistent hash.

**Explanation of the algorithm**

Jump consistent hash works by computing when its output changes as the number of buckets increases. Let ch(key, num_buckets) be the consistent hash for the key when there are num_buckets buckets. Clearly, for any key, k, ch(k, 1) is 0, since there is only the one bucket. In order for the consistent hash function to balanced, ch(k, 2) will have to stay at 0 for half the keys, k, while it will have to jump to 1 for the other half. In general, ch(k, n+1) has to stay the same as ch(k, n) for n/(n+1) of the keys, and jump to n for the other 1/(n+1) of the keys.

Here are examples of the consistent hash values for three keys, k1, k2, and k3, as num_buckets goes up:

|    | 1 | 2 | 3 | 4 | 5 | 6 | 7 | 8 | 9 | 10 | 11 | 12 | 13 | 14 |
|----|---|---|---|---|---|---|---|---|---|----|----|----|----|----|
| k1 | 0 | 0 | 2 | 2 | 4 | 4 | 4 | 4 | 4 | 4  | 4  | 4  | 4  | 4  |
| k2 | 0 | 1 | 1 | 1 | 1 | 1 | 1 | 7 | 7 | 7  | 7  | 7  | 7  | 7  |
| k3 | 0 | 1 | 1 | 1 | 1 | 5 | 5 | 7 | 7 | 7  | 10 | 10 | 10 | 10 |

A linear time algorithm can be defined by using the formula for the probability of ch(key, j) jumping when j increases. It essentially walks across a row of this table. Given a key and number of buckets, the algorithm considers each successive bucket, j, from 1 to num_buckets-1, and uses ch(key, j) to compute ch(key, j+1). At each bucket, j, it decides whether to keep ch(k, j+1) the same as ch(k, j), or to jump its value to j. In order to jump for the right fraction of keys, it uses a pseudo-random number generator with the key as its seed. To jump for 1/(j+1) of keys, it generates a uniform random number between 0.0 and 1.0, and jumps if the value is less than 1/(j+1). At the end of the loop, it has computed ch(k, num_buckets), which is the desired answer. In code:

```
int ch(int key, int num_buckets) {
   random.seed(key);
   int b = 0;  // This will track ch(key, j+1).
   for (int j = 1; j < num_buckets; j++) {
      if (random.next() < 1.0 / (j + 1)) b = j;
   }
   return b;
}
```

We can convert this to a logarithmic time algorithm by exploiting that ch(key, j+1) is usually unchanged as j increases, only jumping occasionally. The algorithm will only compute the destinations of jumps -- the j's for which ch(key, j+1) ≠ ch(key, j). Also notice that for these j's, ch(key, j+1) = j. To develop the algorithm, we will treat ch(key, j) as a random variable, so that we can use the notation for random variables to analyze the fractions of keys for which various propositions are true. That will lead us to a closed form expression for a pseudo-random variable whose value gives the destination of the next jump.

Suppose that the algorithm is tracking the bucket numbers of the jumps for a particular key, k. And suppose that b was the destination of the last jump, that is, ch(k, b) ≠ ch(k, b+1), and ch(k, b+1) = b. Now, we want to find the next jump, the smallest j such that ch(k, j+1) ≠ ch(k, b+1), or equivalently, the largest j such that ch(k, j) = ch(k, b+1). We will make a pseudo-random variable whose value is that j. To get a probabilistic constraint on j, note that for any bucket number, i, we have j ≥ i if and only if the consistent hash hasn't changed by i, that is, if and only if ch(k, i) = ch(k, b+1). Hence, the distribution of j must satisfy

$$P(j \geq i) = P(\text{ch}(k, i) = \text{ch}(k, b+1))$$

Fortunately, it is easy to compute that probability. Notice that since $P(\text{ch}(k, 10) = \text{ch}(k, 11))$ is 10/11, and $P(\text{ch}(k, 11) = \text{ch}(k, 12))$ is 11/12, then $P(\text{ch}(k, 10) = \text{ch}(k, 12))$ is 10/11 * 11/12 = 10/12. In general, if $n \geq m$, $P(\text{ch}(k, n) = \text{ch}(k, m)) = m / n$. Thus for any $i > b$,

$$P(j \geq i) = P(\text{ch}(k, i) = \text{ch}(k, b+1)) = (b+1) / i .$$

Now, we generate a pseudo-random variable, r, (depending on k and j) that is uniformly distributed between 0 and 1. Since we want $P(j \geq i) = (b+1) / i$, we set $P(j \geq i)$ iff $r \leq (b+1) / i$. Solving the inequality for i yields $P(j \geq i)$ iff $i \leq (b+1) / r$. Since i is a lower bound on j, j will equal the largest i for which $P(j \geq i)$, thus the largest i satisfying $i \leq (b+1) / r$. Thus, by the definition of the floor function, $j = \text{floor}((b+1) / r)$.

Using this formula, jump consistent hash finds ch(key, num_buckets) by choosing successive jump destinations until it finds a position at or past num_buckets. It then knows that the previous jump destination is the answer.

```
int ch(int key, int num_buckets) {
   random.seed(key);
   int b = -1;  // bucket number before the previous jump
   int j = 0; // bucket number before the current jump
   while (j < num_buckets) {
      b = j;
      r = random.next();
      j = floor((b + 1) / r);
   }
   return = b;
}
```

To turn this into the actual code of figure 1, we need to implement random. We want it to be fast, and yet to also to have well distributed successive values. We use a 64-bit linear congruential generator; the particular multiplier we use produces random numbers that are especially well distributed in higher dimensions (i.e., when successive random values are used to form tuples) [2]. We use the key as the seed. (For keys that don't fit into 64 bits, a 64 bit hash of the key should be used.) The congruential generator updates the seed on each iteration, and the code derives a double from the current seed. Tests show that this generator has good speed and distribution.

It is worth noting that unlike the algorithm of Karger et al., jump consistent hash does not require the key to be hashed if it is already an integer. This is because jump consistent hash has an embedded pseudorandom number generator that essentially rehashes the key on every iteration. The hash is not especially good (i.e., linear congruential), but since it is applied repeatedly, additional hashing of the input key is not necessary.

**Performance Analysis**

The time complexity of the algorithm is determined by the number of iterations of the while loop. The while loop visits successive jump destinations, which are all less than the number of buckets n except for the last. Thus the expected number of iterations is one more than the the expected number of jumps below n. Since the chance that there is a jump at number of buckets i is 1/i, the expected number of jumps to destinations less than n is just the sum of 1/i for i from 2 to n, which is less than ln(n). So the expected number of iterations is less than ln(n) + 1.

It is interesting to note that jump consistent hash makes fewer expected jumps (by a constant factor) than the log2(n) comparisons needed by a binary search among n sorted keys.

**Performance Measurements**

**Key Distribution**

First we investigate how the algorithms compare in terms of distributing the keys uniformly among buckets. Recall that each key is first mapped to an integer hash value, which is then mapped to a corresponding bucket. The first step is common to both algorithms, so we focus on the second step. Ideally, all buckets should receive the same fraction of hash values. We can measure the deviation from this ideal by computing the standard error ($\sigma/\mu$) of the fraction of hash values assigned to each bucket. Note that for Karger et al.'s algorithm, this depends on the number of points chosen per bucket. The following table summarizes the results:

| Algorithm | Points per Bucket | Standard Error | Bucket Size 99% Confidence Interval |
|---|---:|---:|:---:|
| Karger et al. | 1 | 0.9979060 | (0.005, 5.25) |
|  | 10 | 0.3151810 | (0.37, 1.98) |
|  | 100 | 0.0996996 | (0.76, 1.28) |
|  | 1000 | 0.0315723 | (0.92, 1.09) |
| Jump Consistent Hash |  | 0.00000000764 | (0.99999998, 1.00000002) |

Figure 2: Measures of key space distribution uniformity among buckets.

The last column gives a 99% confidence interval for the bucket size compared to the average bucket size. For example, if Karger et al.'s algorithm is used with 10 points per bucket, then approximately 1% of the buckets will be less than 0.37x smaller than average or more than 1.98x larger than average. This can lead to obvious problems with respect to load balancing. Even with 1000 points per bucket, approximately 1% of buckets will be at least 8% larger or smaller than average. In contrast jump consistent hash divides the key space almost perfectly.

Of course there are also variations in bucket size due to the distribution of the actual keys. For many data storage applications, however, there will typically be millions of keys per bucket (e.g. where each key corresponds to a file, URL, document, etc), in which case the variations due to key distribution are negligible compared with the variations described above.

The key distribution also affects the rebalancing workload when the number of buckets changes. When a new bucket is added with jump consistent hash, the new bucket receives an equal fraction of the key space of each existing bucket. For example, if there are 1000 buckets and a new bucket is added, then each existing bucket will transfer almost exactly 0.1% of its key space. With Karger et al.'s algorithm, on the other hand, the only buckets that participate in rebalancing are the ones that previously contained the points chosen to represent the new bucket. For example, if there are 1000 buckets with 10 points per bucket, then at most 10 buckets will send some of their data to the new bucket. The amount of data from each bucket can also vary substantially. This might cause problems, for example, if a bucket is being added in order to relieve a "hot spot" among the existing buckets: the new bucket has no effect unless the hot spot happens to be one of the 10 buckets selected, but on the other hand if the hot spot is selected, then it may not be able to handle the extra workload of transferring a large fraction of its data. This provides another reason for using a large number of points per bucket with the Karger et al. algorithm.

**Space Requirements**

Distributing the keys uniformly among buckets requires using many points per bucket in Karger et al.'s algorithm, but this increases memory requirements significantly. We implemented two variations of Karger et al.'s algorithm in addition to jump consistent hash. All implementations are in C++ and use the Standard Template Library. They were compiled on a 64-bit platform using Gnu C++ and measured on an Intel Xeon E5-1650 CPU with 32GB of memory.

Our first implementation of Karger et al.'s algorithm ("version A") represents the point data as an STL map from a 64-bit hash value to a 32-bit bucket number. This is probably easiest and most natural way to implement the algorithm. Internally the map is represented as a balanced binary tree. This implementation uses 48 bytes per point per bucket.

The second implementation ("version B") represents the point data as a sorted vector of (hash value, bucket number) pairs, where the hash values are truncated to 32 bits to save space. The bucket corresponding to a given hash value is located using binary search. This implementation uses less space (8 bytes per point per bucket), but unlike the previous implementation, it does not support dynamic updates efficiently: in order to change the number of buckets, the entire data structure must be rebuilt.

The table below presents the total data size for various numbers of buckets, assuming that 1000 points per bucket are used (following the example of Karger et al. in their paper).

| Number of Buckets | Space (Karger, Version A) | Space (Karger, Version B) |
|---:|---:|---:|
| 10 | 469 KB | 78 KB |
| 1000 | 46 MB | 7.6 MB |
| 100000 | 4.5 GB | 0.75 GB |

Figure 3: Space requirements of the Karger et al. implementations.

These relatively large memory requirements are a significant disadvantage when consistent hashing is used to map requests to servers. Any client that wishes to map a request to the correct server must have a copy of the consistent hashing data available locally (or else incur the expense of additional network hops to route the request to its correct destination). This gives jump consistent hash a significant advantage as the number of buckets grows.

**Execution Time**

We measured the execution time of both algorithms on the platform described above, using a benchmark that computes the consistent hash values of a pseudorandom sequence of integer keys.

| Number of buckets | Jump Consistent Hash | Karger A k=10 | Karger A k=100 | Karger A k=1000 | Karger B k=10 | Karger B k=100 | Karger B k=1000 |
|---:|---:|---:|---:|---:|---:|---:|---:|
| 2 | 12 | 25 | 44 | 73 | 26 | 45 | 63 |
| 5 | 20 | 31 | 54 | 92 | 33 | 54 | 70 |
| 20 | 33 | 44 | 73 | 156 | 44 | 65 | 84 |
| 150 | 50 | 68 | 140 | 262 | 62 | 81 | 124 |
| 1024 | 65 | 120 | 231 | 658 | 78 | 114 | 194 |
| 8192 | 81 | 225 | 608 | 1151 | 114 | 185 | 432 |
| 65536 | 96 | 548 | 1088 | 1814 | 188 | 418 | 777 |
| 1048576 | 116 | | | | | | |
| 1073741824 | 165 | | | | | | |

Figure 4: Execution times with no cache competition. Highlighted columns are graphed below.

Figure 4 compares jump consistent hash to the two Karger et al. implementations with k=10, k=100, and k=1000 points per bucket, and the corresponding graph illustrates the k=1000 case only. All times are in nanoseconds and do not include loop overhead. Figure 5 graphs the highlighted columns of figure 4.

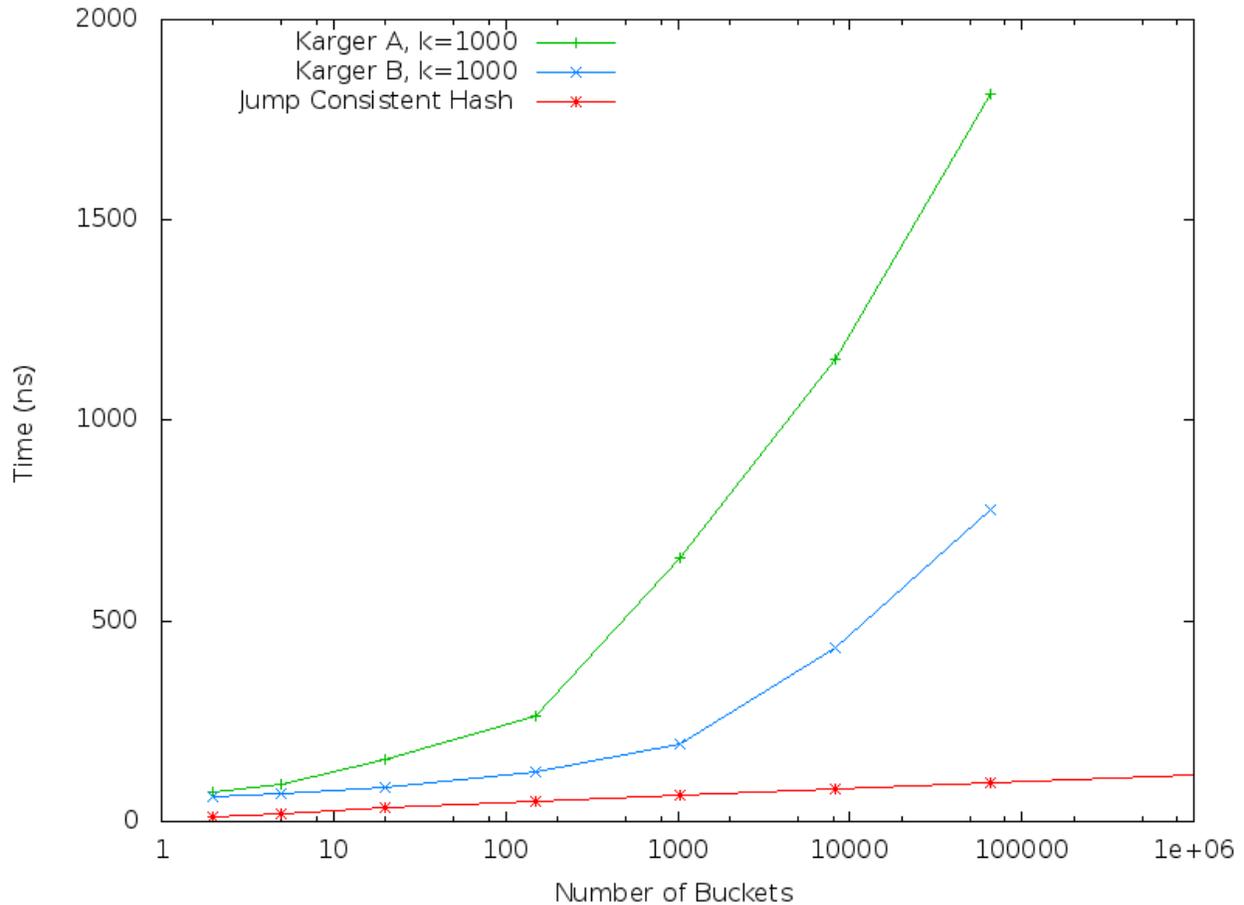

Figure 5: Graph of execution time of jump consistent hash vs. the Karger et al. implementations.

A few points are worth noting:
- The cost of jump consistent hash is logarithmic in the number of buckets, even as the number of buckets grows very large (billions).
- While the Karger et al. implementations also have O(log n) running times, their performance drops off substantially as the data size gets larger due to cache misses (which can increase the running time by a large constant factor).
- With 1000 points per bucket, jump consistent hash is 3-4x faster for up to 1000 buckets, and perhaps 5-8x faster for up to 100,000 buckets.
- Jump consistent hash is still faster than the the Karger et al. implementations even when only 10 points per bucket are used (although this is too few to be practical).

More significantly, the execution times above assume that there is no work being done besides consistent hashing. But real applications typically have much other work that needs to be done as well, which competes for the various levels of memory cache. A typical server might receive a request from somewhere, look up some information in its own data structures, and then send off one or more requests to a data storage system to fetch additional data needed to satisfy the request. This last step uses consistent hashing. But for every consistent hashing calculation, the application almost certainly makes quite a few other memory accesses.

To simulate the behavior of a typical server, we created a benchmark that allocates an additional 1 GB of memory to correspond to the internal data maintained by a server. For each consistent hash calculation, the benchmark reads 16 random bytes within this memory to simulate hash table lookups, pointer following, etc. It also reads one 64K contiguous block within this memory to simulate access to a large in-memory cache.

The following table and graph show the timings of the consistent hash implementations in this environment. As before, all times are in nanoseconds and loop overhead has been subtracted.

| Number of buckets | Jump Consistent Hash | Karger A k=1000 | Karger B k=1000 |
| ---: | ---: | ---: | ---: |
| 2 | 17 | 262 | 72 |
| 5 | 26 | 304 | 86 |
| 20 | 40 | 401 | 115 |
| 150 | 54 | 766 | 187 |
| 1024 | 67 | 1075 | 341 |
| 8192 | 86 | 1540 | 618 |
| 65536 | 103 | 2221 | 966 |
| 1048576 | 121 | | |
| 1073741824 | 176 | | |

Figure 6: Execution times with memory cache competition (simulating a typical server).

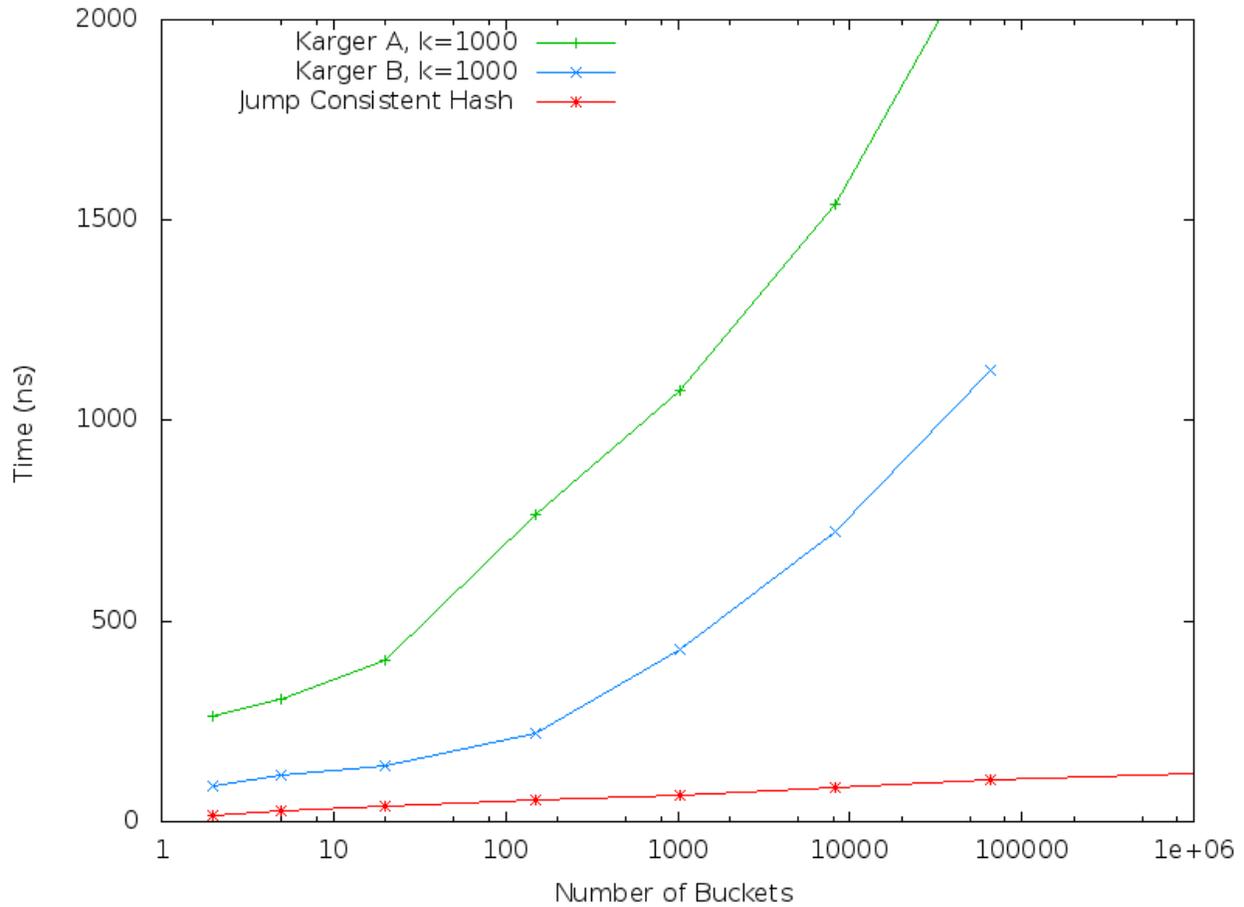

Figure 7: Graph of execution times in the presence of memory cache competition.

The benchmark changes affect jump consistent hash only slightly, but for the Karger implementations the "knee" in the curves is shifted over significantly (i.e., the point where there is a large constant-factor increase in running time due to cache misses). This corresponds to the fact that less cache space is available for consistent hashing because of competition from other code.

**Initialization Time**

The Karger et al. algorithm can also require a significant amount of time to build its data structures. The following table shows the time (in seconds) to build the data structures for various numbers of buckets with k=1000 points per bucket.

| Number of Buckets | Karger A, k=1000 | Karger B, k=1000 |
|---|---|---|
| 2 | 0.00024 | 0.00011 |
| 5 | 0.00072 | 0.00031 |
| 20 | 0.0039 | 0.0014 |
| 150 | 0.045 | 0.012 |
| 1024 | 0.61 | 0.093 |
| 8192 | 8.94 | 0.85 |
| 65536 | 111.99 | 7.66 |

The main points worth noting are that the balanced-tree implementation (STL map) is relatively slow to initialize, and that both implementations have very significant initialization times as the number of buckets grows large. Also note that with the Karger B implementation, the data structure must be completely rebuilt whenever the number of buckets changes.

**Acknowledgements**

Chad Lester contributed to discussions on how to reshard large data storage systems. Eric Lehman simplified the derivation of the performance bound.

**Note**

Google has not applied for patent protection for this algorithm, and, as of this writing, has no plans to. Rather, it wishes to contribute this algorithm to the community.